\documentclass[aps, prl, twocolumn, groupedaddress, 10pt, showpacs ,superscriptaddress]{revtex4-1}

\usepackage{graphicx}
\usepackage{color}
\usepackage{amsfonts, amsmath, amsthm,amssymb, amscd}

\usepackage{hyperref}

\def\epp{\: .}
\def\epc{\: ,}

\def\blam{{\boldsymbol{\lambda}}}
\def\bmu{{\boldsymbol{\mu}}}

\def\asp{a_\text{sp}}

\def\rhosp{\rho_\text{sp}}

\begin{document}

\title{Relaxation dynamics of local observables in integrable systems}

\author{J. De Nardis}
\affiliation{Institute for Theoretical Physics, University of Amsterdam, Science Park 904\\
Postbus 94485, 1090 GL Amsterdam, The Netherlands}

\author{L. Piroli}
\affiliation{International School for Advanced Studies (SISSA) and INFN,\\
via Bonomea 265, 34136 Trieste, Italy}

\author{J.-S. Caux}
\affiliation{Institute for Theoretical Physics, University of Amsterdam, Science Park 904\\
Postbus 94485, 1090 GL Amsterdam, The Netherlands}

\date{\today}

\begin{abstract}
We show, using the quench action approach \cite{2013_Caux_PRL_110}, that the whole post-quench time evolution of an integrable system in the thermodynamic limit can be computed with a minimal set of data which are encoded in what we denote the generalized single-particle overlap coefficient $s_0^{\Psi_0}(\lambda)$. This function can be extracted from the thermodynamically leading part of the overlaps between the eigenstates of the model and the initial state. For a generic global quench the shape of $s_0^{\Psi_0}(\lambda)$ in the low momentum limit directly gives the exponent for the power law decay to the effective steady state.  As an example we compute the time evolution of the static density-density correlation in the interacting Lieb-Liniger gas after a quench from a Bose-Einstein condensate. This shows an approach to equilibrium with power law $t^{-3}$ which turns out to be independent of the post-quench interaction and of the considered observable.    
\end{abstract}


\maketitle

\paragraph*{Introduction.}
Understanding the non-equilibrium time evolution of a many-body interacting system is one of the main challenges in contemporary physics \cite{2011_Polkovnikov_RMP_83,2015_Eisert}. The study of systems with nontrivial interactions among their constituents is hard enough when the system is in its ground state; things however get even more complicated out of equilibrium, since most of the usual theoretical tools then become inapplicable. This is mainly due to the high energy regions of the spectrum which are probed by the time evolution, where the mean field approach and low energy approximations are not valid. Numerical simulations on the other hand are severely limited in the range of time or of system sizes \cite{DMRG}. Developing new methods able to predict the short-, intermediate- and long-time dynamics of an out-of-equilibrium system is thus an urgent priority, especially in view of the rapid progress achieved in experiments on ultracold atoms \cite{2006_Kinoshita_NATURE_440,2007_Hofferberth_NATURE_449,2012_Trotzky_NATPHYS_8,
2012_Cheneau_NATURE_481,2012_Gring_SCIENCE_337,2013_Meinert_PRL,2014_Meinert_SCIENCE_344,2015_Langen_SCIENCE_348}. 

Since the beginnings of quantum mechanics \cite{Neumann_1929}, much interest has been devoted to the fundamental problem of calculating the time dependence of physical observables in states which are not eigenstates of the Hamiltonian driving the time evolution. This situation has now come to be known as a quantum quench \cite{2006_Calabrese_PRL_96,2011_Polkovnikov_RMP_83} and has been of major interest both from experimental 
 and theoretical points of view. Most of the theoretical research focused so far on the expectation values of local observables at late times after the quench, when the system is in an effective steady state. In particular the Generalized Gibbs Ensemble (GGE) hypothesis \cite{2007_Rigol_PRL_98,2008_Rigol_NATURE_452} focuses on the possibility of reducing the huge complexity of the initial wave function to a reduced set of information, incorporated in the local conserved quantities of the system, which gives the expectation values of all physical observables in the steady state. However the question of how to perform an analogous simplification for the whole post-quench time evolution, much more relevant from the experimental point of view, has been poorly addressed, except in a few cases \cite{2006_Calabrese_PRL_96,2010_Fioretto_NJP_12,2010_Gritsev_JSTAT_P05012,2011_Calabrese_PRL_106,2012_Calabrese_JSTAT_P07016,
 2012_Mossel_NJP_7,2012_Karrasch_PRL_109,2013_Caux_PRL_110,2013_Collura_PRL_110,2013_Faribault_PRL_110,2013_Mussardo_PRL_111,2014_Kormos_PRA_89,Essler_PRL_2014,2014_Fioretto_NJP,2014_Bertini_Sinegordon,Sotiriadis201452,2014_Sotiriadis_JSTAT_P07024,2014_Essler_PRB_89,2013_Marcuzzi_PRL,2013_Fagotti_PRB_87,2014_Fagotti_PRB_89,2014_DeNardis_JSTAT_12,2015_Bertini_arXiv}.

The quench action method introduced in \cite{2013_Caux_PRL_110} has recently proved to offer a procedure whereby one can derive, from first principles, not only the steady state itself but also the actual time evolution of physical observables  \cite{2013_Caux_PRL_110,2014_DeNardis_PRA_89,2014_Wouters_PRL_113,2014_Brockmann_JSTAT_P12009,2014_Pozsgay_Dimer,2014_Bertini_Sinegordon,2014_DeNardis_JSTAT_12,2015_De_Luca_PRA,2015_Bettelheim_JPHYSA}. In summary, this method combines knowledge of initial state overlaps with functional integration techniques to extract the thermodynamically relevant information on the relaxation dynamics of an integrable system. The purpose of this Letter is to show that this approach, combined with the recent observations on the structure of the overlaps between eigenstates of different Hamiltonians \cite{2014_DeNardis_PRA_89,2014_Brockmann_JPA,2014_Brockmann_npi,2014_Brockmann_LL_qq,2014_Piroli_resursive_overlaps}, is able to provide the full post-quench time evolution in terms of a reduced set of data which can be extracted from the thermodynamically leading part of the overlaps. It turns out that the same function, the generalized single-particle overlap coefficient $s_0^{\Psi_0}(\lambda)$, fixes the steady-state expectation values and the whole time evolution from $t=0^+$ after the quench. This is treated analogously to a system at thermal equilibrium with a sub-entropic gas of independent particle-hole excitations around the steady state constituting the whole effective spectrum necessary to compute the time dependence of all physical observables. A restricted class of excitations is then clearly seen to be the most relevant for the long-time behavior, giving a picture reminiscent of a field theory description of the asymptotics of correlations in equilibrium situations \cite{2009_Imambekov_SCIENCE_323,2009_Imambekov_PRL_102,2014_Karrasch}.\\

This letter is organized as follows. First we show how, for a generic integrable model, the quench action method \cite{2013_Caux_PRL_110} allows to extract the whole post-quench time evolution from the complex function $s_0^{\Psi_0}(\lambda)$ denoted here as the generalized single-particle overlap coefficient. Then we specialize to the time evolution of the static density moment $g_2(x=0,t)$ of the interacting Lieb-Liniger gas after a quench from the ground state of the bosonic free theory. The same quench has been studied in a number of recent works \cite{2010_Gritsev_JSTAT_P05012,2012_Deepak_PRL_109,2013_Kormos_PRB_88,2013_Mussardo_PRL_111,2014_Mazza_JSTAT,2014_Kormos_PRB_89,2014_Essler_Panfil,2015_Zill_PRA}. We use here the exact results for the post-quench saddle point reported in \cite{2014_DeNardis_PRA_89} and we obtain a rare full post-quench time evolution of a physical observable in a truly interacting model that is closely related to recent experiments \cite{2005_Kinoshita_PRL_95,2015_Fabbri_PRA}.

\paragraph*{Time evolution in an integrable model}
We consider an initial state $| \Psi_0 \rangle$ which is not an eigenstate of the one-dimensional integrable Hamiltonian $H$ for $N$ particles moving on a system size $L$ with periodic boundary conditions. In a generic integrable model each eigenstate is specified by a set of $N$ quantum numbers $\boldsymbol{I} = \{ I_j\}_{j=1}^N$. The set of nonlinear coupled Bethe equations maps the mutually excluding quantum numbers in $N$ quasi-momenta, called rapidites,  $\boldsymbol{I} \to \boldsymbol{\lambda} = \{ \lambda_j\}_{j=1}^N$ which take value in the complex plane. These are related to the one-particle momentum $k_0(\lambda)$ and the scattering phase of the model $\theta(\lambda)$ \cite{KorepinBOOK}
\begin{equation}\label{eq:BE}
k_0(\lambda_i) = \frac{2 \pi I_i}{L}  - \sum_{k=1}^N \theta(\lambda_i - \lambda_k) \qquad i = 1, \ldots, N \epp
\end{equation}
All the possible different choices of quantum numbers $\boldsymbol{I}$ give a complete basis of eigenstates $| \blam\rangle$ with energy $E[\blam]$ which allows, given a local operator ${\hat{O}}$, to resolve the time evolution of its expectation value on the initial state $\langle  \Psi_0 | e^{i H t}  \hat{O} e^{- i H t} | \Psi_0 \rangle  \equiv \langle \hat{O}(t) \rangle $
\begin{align}\label{eq:time_ev_exact}
 \langle \hat{O}(t) \rangle  = \sum_{\blam}  \sum_{\bmu} e^{- S^{\Psi_0}_{\blam}} e^{- (S^{\Psi_0})^*_{\bmu}}\langle \blam | \hat{O} | \bmu \rangle e^{- i t (E [\bmu] - E[\blam])} \epc
\end{align}
where we introduced the overlap coefficients $S^{\Psi_0}_{\blam}$ between the initial state and the eigenstates $\langle \Psi_0 | \blam \rangle =  e^{- S^{\Psi_0}_{\blam}}$. 
The double sum in \eqref{eq:time_ev_exact} can be performed in general when the number $N$ of constituents of the system is small. However one is in general interested in the thermodynamic limit $\lim_{\text{th}} \equiv \lim_{N,L \to \infty}$ with fixed density $n=N/L$. The quench action approach introduced in \cite{2013_Caux_PRL_110} allows to move from a sum over the discrete representation for the eigenstates, in terms of the quantum numbers $\boldsymbol{I}$ to a functional integral over smooth distributions of rapidites and simple excitations over them. Given a smooth function  $\rho(\lambda)$  of rapidities on the real axis with its normalization given by the density of particles $\int_{-\infty}^{\infty} d\lambda \: \rho(\lambda) = n$ (under the string hypothesis it can be generalized to complex rapidities \cite{TakahashiBOOK}), there is an entropic number $\sim e^{S_{YY}[\rho]}$ of finite size states that share the same expectation values of local operators \cite{KorepinBOOK}.  The entropy is given by \cite{1969_Yang_JMP_10} 
\begin{equation}
S_{YY}[\rho] =  L \int_{-\infty}^{\infty} d\lambda \big( (\rho +  \rho^h) \ln (\rho + \rho^h) -   \rho \ln \rho
  - \rho^h \ln \rho^h   \big) \epc
\end{equation}
where the density of holes is given by the total density $\rho^h = \rho^t - \rho$, related to the density of particles by the Bethe equations \eqref{eq:BE} in the thermodynamic limit
\begin{equation}
\rho^t  = \frac{1}{2 \pi} \left( \dot{k}_0 +    \dot{\theta} \ast \rho  \right) \epc
\end{equation} 
where we introduce the convolution between two functions $f \ast g = \int_{-\infty}^{\infty} d\mu f(\lambda - \mu) g(\mu)$ and the derivative respect to $\lambda$, $\frac{d f}{d \lambda} \equiv  \dot{f}$. For later convenience we also introduce the scalar product on the real axis $f \cdot g =  \int_{-\infty}^{\infty} d\mu f(\mu) g(\mu)$. After restricting to the appropriate sub-Hilbert space with nonzero overlap (if discrete symmetries are present), the overlaps become a smooth functional over the eigenstates. In particular they can be written as an extensive universal part (dependent only on the distribution $\rho$) with subleading corrections which depend on the finite number of particle-hole excitations $\{ h_i , p_i \}_{i=1}^m$ over the distribution $\rho$ (which corresponds to displacing a number $m$ of quantum numbers of one of the finite size state $| \blam \rangle \to | \rho \rangle $ which discretizes the distribution $\rho$)
\begin{equation}
S^{\Psi_0}_{\blam} \to \mathcal{S}[\rho] + \delta s[\rho, \{ h_i , p_i\}_{i=1}^m] + \mathcal{O}(1/N) \epc
\end{equation}
where both quantities are given in terms of the generalized one-particle overlap coefficient $s_0^{\Psi_0}(\lambda) $
\begin{align}\label{eq:overlaps1}
 \mathcal{S}[\rho] & = L \:  s_0^{\Psi_0} \cdot \rho  \nonumber \epc\\ 
 \delta s[\rho, \{ h_i,p_i\}_{i=1}^m] & = \sum_{k=1}^m \Big(  
s_0^{\Psi_0}(p_k) -  s_0^{\Psi_0}(h_k) -  F_k \cdot \dot{s}_0^{\Psi_0} \Big) \epp
\end{align}
The back-flow $F_k(\lambda)$ for a single particle-hole is computed in terms of the distribution $\rho$
\begin{align}
2 \pi F_k \frac{\rho^t}{\rho} = &\theta(\lambda - p_k) -  \theta(\lambda - h_k)  + \dot{\theta} \ast F_k \epp
\end{align}
Therefore in the thermodynamic limit, for any \textit{weak} operator \footnote{We use \textit{weak} here to denote observables which do not reorganize the steady state, in other words which are not entropy-producing. This class includes all the local observables},  we can write its time-dependent expectation value \eqref{eq:time_ev_exact} as
\begin{align}\label{eq:sumsum}
 & \lim_{\text{th}} \langle \hat{O}(t) \rangle   = \frac{1}{2} {\int D\rho  \: e^{- 2 \Re \mathcal{S}[\rho] +   S_{YY}[\rho]}  } \sum_{m=0}^\infty \int d[h,p]_m  \nonumber \\   &  \Big[ e^{-  \delta s[\rho, \{ h_i , p_i\}_{i=1}^m] - i t \omega[\rho, \{ h_i , p_i\}_{i=1}^m]} \langle \rho |  \hat{O} | \rho, \{ h_i , p_i \}_{i=1}^m \rangle \Big]\nonumber \\& + \text{mirr} \epc
\end{align}
with $\int d[h,p]_m  = \frac{1}{(m!)^2} \prod_{j=1}^m \int_{-\infty}^{\infty} dh_j \rho(h_j) \int_{-\infty}^{\infty} dp_j \rho^h(p_j)$ denoting the sum over the macroscopic particle-hole excitations and $\text{mirr}$ indicating the same sum as in \eqref{eq:sumsum} but with excitations on the left state \footnote{It corresponds to the complex conjugate of \eqref{eq:sumsum} when $ \hat{O}$ is a Hermitian operator}. The energy of a state $E[{\blam}] \to \mathcal{E}[\rho] +  \omega [\rho, \{ h_i , p_i\}_{i=1}^m]$ is given in terms of the one-particle energy $\epsilon_0(\lambda)$ analogously to the overlaps \eqref{eq:overlaps1}
\begin{align}
 \mathcal{E}[\rho] & = L \:  \epsilon_0 \cdot \rho \nonumber \epc \\
  \omega[\rho, \{ h_i,p_i\}_{i=1}^m] & = \sum_{k=1}^m \Big(  
\epsilon_0(p_k) -  \epsilon_0(h_k) \nonumber-  F_k \cdot \dot{\epsilon}_0 \Big) \epp
\end{align}
The matrix elements $\langle \rho |  \hat{O} | \rho \rangle $ can be computed by choosing one of the possible (large) finite size realizations $ | \blam \rangle \to | \rho \rangle $ of the distribution $\rho(\lambda)$ and using $ \frac{\langle \rho | \hat{O} | \rho \rangle }{\langle \rho | \rho \rangle} = \frac{\langle \blam | \hat{O} | \blam \rangle}{\langle \blam | \blam \rangle}\Big( 1 + \mathcal{O}(1/N)\Big)$. The same can be done for the off diagonal ones. 
Given these ingredients the sum in \eqref{eq:time_ev_exact} can be evaluated in the saddle point $\frac{\delta S^{Q}[\rho] }{\delta \rho} \Big|_{\rho = \rho_{sp}} =0$ of the quench action 
$S^{Q}[\rho] = 2 \Re \mathcal{S}[\rho] - S_{YY}[\rho]$
leading to an expression for the whole post-quench time evolution in the thermodynamic limit \cite{2013_Caux_PRL_110,2014_DeNardis_PRA_89}
\begin{align}\label{eq:time_ev_QA}
\lim_{\text{th}} & \langle \hat{O}(t) \rangle   = 
\frac{1}{2} \sum_{m=0}^{\infty}\int d[h,p]_m  \Big[ e^{-i t (\omega[\rho_{sp}, \{ h_i , p_i\}_{i=1}^m]) }   \nonumber \nonumber \\ &  \times  e^ {  -  \delta s[\rho_{sp}, \{ h_i , p_i\}_{i=1}^m] }\langle \rho_{sp} | \hat{O} | \rho_{sp} , \{h_i , p_i\}_{i=1}^m \rangle \Big] + \text{mirr.} 
\end{align}
A notable consequence of formula \eqref{eq:time_ev_QA} is that all the information to reconstruct the entire post-quench time evolution is contained in the function $s_0^{\Psi_0}(\lambda)$ which can be extracted by taking the scaling limit of the overlap coefficients 
\begin{equation}
S^{\Psi_0}_\blam = \sum_{j=1}^N \left( s_0^{\Psi_0}(\lambda_j) + \mathcal{O}(N^{-1}) \right) \epp
\end{equation}
The behavior of the exponent of $s_0^{\Psi_0}(\lambda)$ around $\lambda \sim 0$ determines the power law for the large time relaxation of any  physical observable. In the limit of large $t$ we can indeed approximate the sum in \eqref{eq:time_ev_QA} with the contribution of the saddle point itself and of the single particle-hole excitations
\begin{align}
& \lim_{\text{th}} \langle \hat{O}(t) \rangle  \sim  \langle \rho_{sp} | \hat{O} | \rho_{sp} \rangle + \frac{1}{2}\int_{-\infty}^{\infty} \: dp \: dh \: \rho_{sp}^h(p)      \rho_{sp}(h)  \nonumber \\ 
 &  \times   \langle \rho_{sp} | \hat{O} | \rho_{sp}  , \{h , p\} \rangle e^{-i t \omega[\rho_{sp}, \{ h , p\}] -  \delta s[\rho_{sp},\{ h , p\}] } + \text{mirr.} 
\end{align}
Since the dispersion relation, as well as the differential overlap coefficient $\delta s[\rho, \{ h , p\}]$, splits in terms of particles and holes $\omega[\rho,\{ h , p\}] \equiv \omega[\rho, p] - \omega[\rho, h]$, the integrals can be approximated by evaluating each of them in the saddle point of the single-particle dispersion relation $\omega[\rho, \lambda]$ which for any smooth distribution $\rho(\lambda)$ is in $\lambda=0$. Therefore if $k$ is the order of the first non-zero derivative in $\lambda=0$ of $e^{ s^{\Psi_0}(\lambda)}$, the approach to the steady state value of all the local operators with a finite expectation value on the saddle point state ($\langle \rho_{sp}|  \hat{O} | \rho_{sp}\rangle \neq 0$) is given by a power law as follows
\begin{align}\label{eq:power_law}
& \Delta_{\hat{O}}(t) \sim t^{- (k+2)} \: \forall k \text{ odd  }  \epc \:\:\:  \Delta_{\hat{O}}(t) \sim t^{- (k+1)}
\:   \forall k \text{ even  } \epc
\end{align}
where $\Delta_{\hat{O}}(t) \equiv  \langle \hat{O} (t)\rangle -  \lim_{{t\to \infty} } \langle   \hat{O}(t) \rangle $. 
The power law decay of correlations is a consequence of the creation, by the quench, of a finite density of holes around $\lambda = 0$, giving a finite density of states for small-energy (zero velocity) particle-hole excitations in this region \cite{2014_Karrasch} (See figure \ref{fig:Fig2}, panel (b)). Therefore the contribution of the power law is proportional to the density of holes around $\lambda \sim 0$ in the post-quench saddle point state which is large for distributions with a large (extensive) entropy. Any initial state with an extensive amount of energy $\langle \Psi_0 | H | \Psi_0 \rangle \sim  L e_0$ shows therefore a relaxation as power law although its contribution to the whole time evolution becomes less and less visible as $e_0$ decreases. Note that up to now we assumed that the operator $\hat{O}$ conserves the total number of particles. For operators adding (or removing) one extra particle to the system the power law is simply replaced by  $t^{- (k+2)/2}$ ($t^{- (k+1)/2}$ for $k$ even).

Finally it is important to note that the same large time decay is expected also for systems with bound states (under the string hypothesis) \cite{2014_Wouters_PRL_113,2014_Brockmann_JSTAT_P12009,2014_Pozsgay_Dimer}. However in this case the full time evolution from $t=0^+$ can only be recovered by including other classes of high energy excitations, namely recombinations between bound states of different masses (strings of different lengths \cite{TakahashiBOOK}). 

\begin{figure}
\begin{center}
\includegraphics[width=0.98\columnwidth]{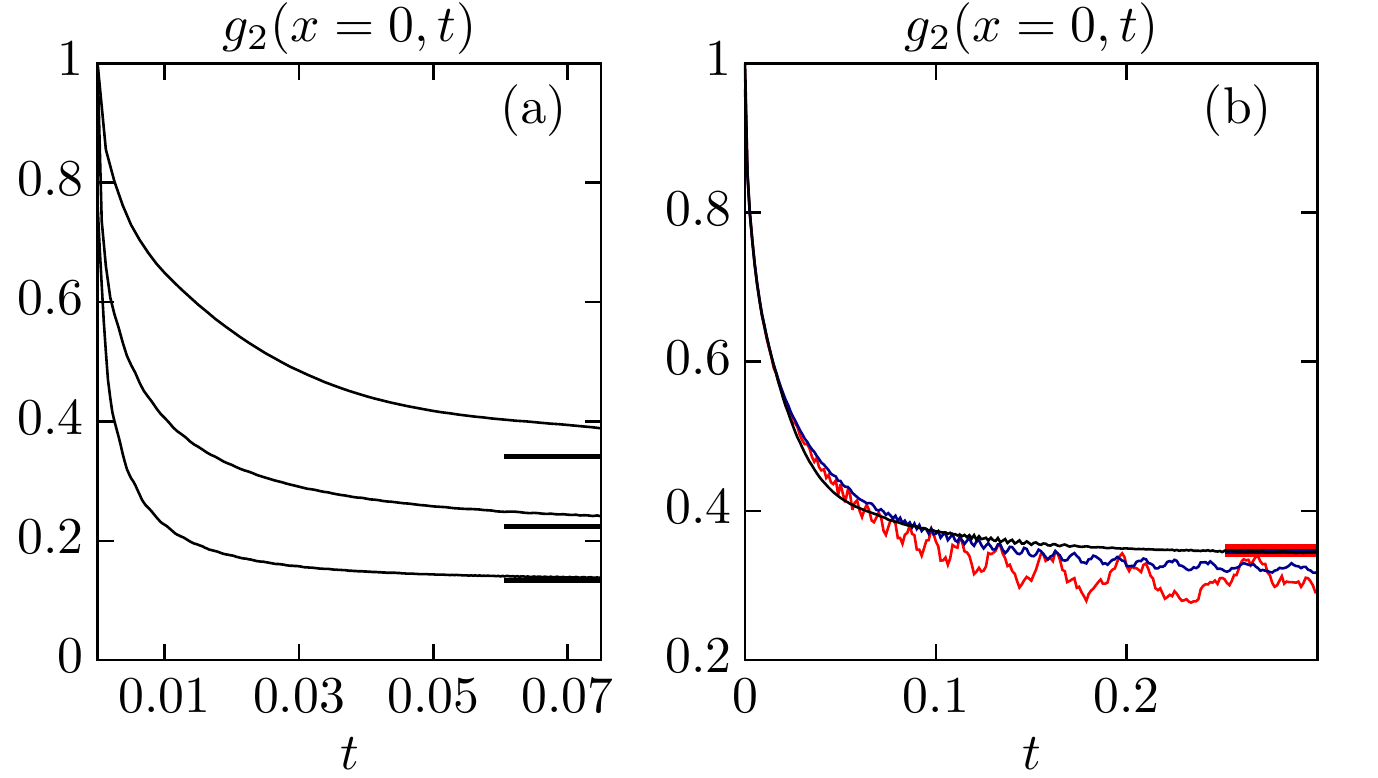}
\caption{\label{fig:Fig1}(a): (color online) Time evolution of $g_2(x=0,t)$ as a function of time for different values of the post-quench interaction $\gamma = 4,8,16$, (from top to bottom) in the thermodynamic limit with fixed density $n=1$. The data are obtained using equation \eqref{eq:time_ev_QA} and by averaging over 10 finite size realizations of the saddle-point states with a system size of $N=96$ particles. The lines on the right respectively indicate the steady state values in the thermodynamic limit as given in \cite{2014_DeNardis_PRA_89}. (b): (color online) Time evolution of $g_2(x=0)$ as a function of time for $\gamma = 4$ and different system sizes: $N=6,8$ (red, blue line) and in the thermodynamic limit (black line). The finite size asymptotic values (red and blue lines on the right) are shown. The data for $N=6$ and $N=8$ are obtained by performing the full double sum over the Hilbert space \eqref{eq:time_ev_exact} and dividing by their initial value $g_2(x=0,t=0) = (1-\frac{1}{N})$. Their asymptotic values correspond to the value of the diagonal ensemble $\frac{\sum_{\blam}   e^{- 2 \Re S^{\Psi_0}_{\blam}} \langle \blam |  :\!(\hat{\rho}(0)/n)^2\!\!:  | \blam \rangle}{g_2(x=0,t=0) } $.}
\end{center}
\end{figure}

\begin{figure}
\begin{center}
\includegraphics[width=0.98\columnwidth]{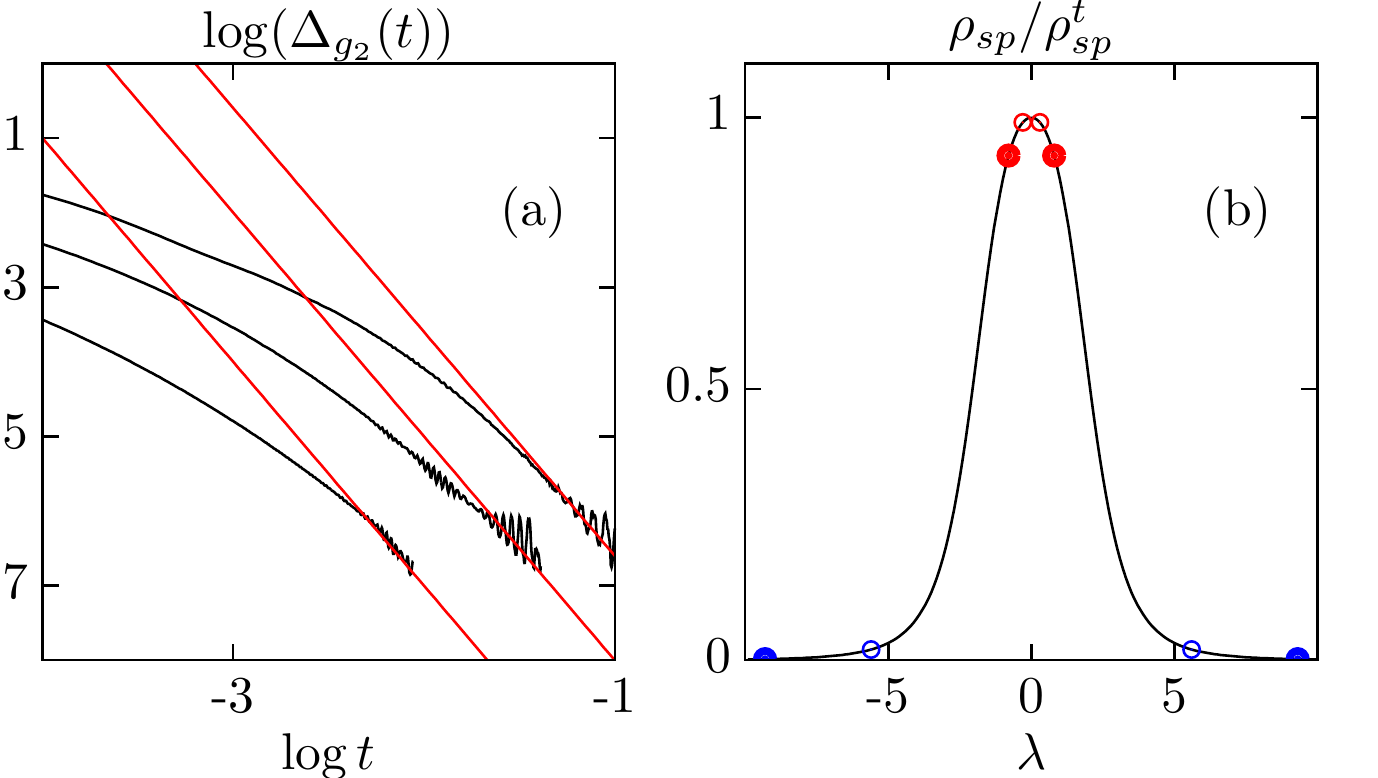}
\caption{\label{fig:Fig2} (a): (color online) Log-log plot of the time evolution of $\Delta_{g_2}(t)$ for different values of the post quench interaction $\gamma=4,8,16$ (from top to bottom). The red lines $f(t)=  \text{const}(\gamma)- 3\log t$ are guide for the eyes showing the approach to the equilibrium value as $\sim t^{-3}$ for all the considered values of the post-quench interactions. (b): (color online) Schematic representations of the most relevant parity invariant particle (filled dots) with its respective hole (empty dot) excitations on the saddle point filling function (here for $\gamma=4$). For small times the high energy excitations corresponding to particle-holes along the tails (blue dots) need to be included. For large times on the other hand the relevant excitations are particle-holes deep in the center of the distribution $\lambda \sim 0$ (red dots) which are the ones responsible for the power law decay of the correlations towards their steady state values.  }
\end{center}
\end{figure}
\paragraph*{Time evolution in the interacting Bose gas}
As a specific example of the general method, we now focus on the Lieb-Liniger model for a $\delta-$interacting Bose gas, defined by the Hamiltonian \cite{1963_Lieb_PR_130_1} (setting $\hbar = 2m = 1$)
\begin{equation}\label{eq:LL_Ham}
H_{LL}=  -\sum_{j=1}^N \frac{\partial^2 }{\partial x^2_j} + 2c \sum_{j>k} \delta(x_j - x_k) \epp
\end{equation}
The initial state is chosen to be the ground state in the absence of interactions $\gamma_0 =0$, where $\gamma = c/n$ effectively parametrizes the coupling in the thermodynamic limit. This state is known as the Bose-Einstein condensate (BEC) state $| \text{BEC}\rangle$ and it is spatially structureless in all coordinates, $\langle \mathbf{x} | \text{BEC} \rangle = \frac{1}{L^{N/2}}$.
We consider the post-quench time evolution of the static density moment $g_2(x=0)$, measuring the rate of two-body inelastic processes in the gas \cite{2003_Gangardt_PRL} which can be experimentally accessed through the measurement of the photoassociation rate \cite{2005_Kinoshita_PRL_95}
\begin{equation}
g_2(x=0,t) = \langle  \text{BEC}| e^{i H_{LL} t} :\!(\hat{\rho}(0)/n)^2\!\!: e^{- i H_{LL} t} |  \text{BEC} \rangle\epp
\end{equation}
The density operator is defined as $\hat{\rho}(x) =\mathbf{\Psi}^{\dagger}(x)\mathbf{\Psi}(x)$, where the bosonic operators $\mathbf{\Psi}(x)$, $\mathbf{\Psi}^{\dagger}(x)$ satisfy the canonical commutation relations $[\mathbf{\Psi}(x), \mathbf{\Psi}^{\dagger} (x')]= \delta(x-x')$. The overlaps and in particular the generalized one-particle overlap coefficient have been computed in \cite{2014_DeNardis_PRA_89}
\begin{equation}\label{eq:s0BEC}
s_0^{\text{BEC}}(\lambda) = \log \left( \frac{\lambda}{c} \sqrt{\frac{\lambda^2}{c^2} + \frac{1}{4}} \right) \epc
\end{equation}
where the branch-cut of the logarithm is chosen such that $s_0^{\text{BEC}}(-\lambda) = - s_0^{\text{BEC}}(\lambda)$.   The one-particle energy and momentum are given by $ k_0 (\lambda)= \lambda$ and $ \epsilon_0(\lambda) = \lambda^2 $.
The function $s_0^{\text{BEC}}(\lambda)$ determines the saddle point state which can be analytically written in terms of Bessel functions of the first kind $I_n(z)$ \cite{2014_DeNardis_PRA_89}
\begin{align}\label{eq:rho}
\rhosp(\lambda) &=-  \frac{\gamma}{4\pi} \frac{1}{1+\asp(\lambda)}   \frac{\partial  \asp(\lambda)}{\partial \gamma} \epc \\
\asp(\lambda) &= \frac{2\pi/\gamma}{\frac{\lambda}{c}\sinh \left(  \frac{2\pi \lambda}{c} \right)} I_{1-2i\frac{\lambda}{c}} \!\! \left( \frac{4}{\sqrt{\gamma}}\, \right) I_{1 + 2i\frac{\lambda}{c}} \!\! \left(  \frac{4}{\sqrt{\gamma}}\, \right) \epp \nonumber 
\end{align}
The matrix elements between the eigenstates of the model are given in \cite{Pozsgay_Local,Piroli}. The sum is performed by averaging over different finite size realizations $| \blam_{sp} \rangle \to | \rho_{sp} \rangle$ of the saddle point state and evaluating the relevant excitations via an adaptation of the ABACUS algorithm \cite{2009_Caux_JMP_50,2006_Caux_PRA_74,2014_Panfil_PRA} to generic highly-excited states. In figure \ref{fig:Fig1} the time evolution computed via the quench action approach \eqref{eq:time_ev_QA} shows that even for values of the coupling constant that are far from the two perturbative regimes (weak and strong coupling) we recover the initial BEC value of the correlation $\lim_{\text{th}} g_2(x=0,t=0^+)  \equiv g_2(x=0)_{BEC}= 1$ \cite{2003_Gangardt_PRL}. The thermodynamic results allow to extract their large time decay to their steady state values as in figure \ref{fig:Fig2}. This follows the expected $t^{-3}$ law which is a consequence of \eqref{eq:power_law} and of the behavior of  $s_0^{\text{BEC}} (\lambda)$ around $\lambda=0$
\begin{align}\label{eq:behaviour_BEC}
& \exp\left(s_0^{\text{BEC}} (\lambda=0) \right)=0  \qquad \forall \: \gamma >0  \epc \\
& \frac{d \exp\left( s_0^{\text{BEC}} (\lambda) \right)}{d \lambda} \Big|_{\lambda=0} \neq 0 \qquad \forall \: \gamma >0   \epp
\end{align}
This shows that the relaxation following a power law is present for any post-quench coupling constant $\gamma$, even in the limit of small interactions. This is in contrast to the predictions of the Bogoliubov approximation where the decay is predicted to be exponential for small $\gamma$ \cite{2010_Carusotto_EPJD_56}. Note that the behavior of the overlap as in \eqref{eq:behaviour_BEC} is also independent of the initial value of the coupling constant. It is related to the fact that for quenches from the ground state of the theory with a coupling $\gamma_0>0$ to the gas with a finite coupling $\gamma >0$ the eigenstate with the maximal overlap $e^{- S^{\Psi_0}_{\blam}}$ is clearly the ground state of the final theory. This leads to the divergent behavior of the generalized single-particle overlap for small values of the rapidity, $ \lim_{\lambda \to 0} e^{-s_0^{\Psi_0}(\lambda) } \to \infty$ which leads to \eqref{eq:behaviour_BEC}. Therefore the same power law $t^{-3}$ is expected for any interaction quench $\gamma_0 \to \gamma>0$ inside the repulsive regime of the one-dimensional Bose gas (for quenches to the free bosonic theory $\gamma_0 \to \gamma =0$ see \cite{2012_Mossel_NJP_7,2014_Sotiriadis_JSTAT_P07024}). 
\paragraph*{Conclusions}
We showed how the quench action approach allows to reconstruct the whole post-quench time evolution of an integrable system from data contained in the thermodynamically leading part of the overlaps. In particular we presented an argument to predict the power law behavior for the late times approach to equilibrium of local observables. This is a direct consequence of the creation in the gas of macroscopic excitations with vanishing velocity which is a generic feature of the model itself, independently of the quench protocol. The question if an adaptation of the non-linear Luttinger liquid approach for equilibrium correlations \cite{2009_Imambekov_SCIENCE_323,2009_Imambekov_PRL_102,2014_Karrasch} can be implemented to compute the late time dynamics after a quench will be addressed in forthcoming works. \\
As a proof of principle we computed the time evolution in the Lieb-Liniger model of the static density moment $g_2(x=0,t)$ after a quench from the Bose-Einstein condensate. This represents a rare example of a full post-quench time evolution of a truly interacting model and therefore it can be directly connected to experimental results in ring-like geometries \cite{2013_Wright_PRA}, box-like potentials \cite{2010_Amerongen} or any other experimental realization of the one-dimensional Bose gas where the confining trap influences time scales which are much larger than the relaxation time of one-point functions as $g_2(x=0,t)$ \cite{2012_Gring_SCIENCE_337,2015_Langen_SCIENCE_348}. The comparison between the finite size calculations and the thermodynamic limit in figure \ref{fig:Fig1} shows indeed that for short times the relaxation processes are well approximated by $N \sim 10$ particles.  This also underlines the importance of obtaining exact results in the thermodynamic limit, that can be used to test numerical simulations for small system sizes as done in \cite{2015_Zill_PRA}. The method can be extended to two-point functions as the dynamical density-density correlations of the gas and to other models as the XXZ spin chain \cite{2014_Wouters_PRL_113,2014_Pozsgay_Dimer}.

\paragraph*{Acknowledgments}
We acknowledge useful and inspiring discussions with F. H. L. Essler, P. Calabrese, G. Mussardo  and M. Panfil. 
We acknowledge support from the Foundation for Fundamental Research on Matter (FOM) and the Netherlands Organisation for Scientific Research (NWO). This work forms part of the activities of the Delta Institute for Theoretical Physics (D-ITP).

\bibliography{G2_biblio}

\end{document}